\documentclass[fleqn,usenatbib]{mnras}

\usepackage{newtxtext,newtxmath}

\usepackage[T1]{fontenc}

\DeclareRobustCommand{\VAN}[3]{#2}
\let\VANthebibliography\thebibliography
\def\thebibliography{\DeclareRobustCommand{\VAN}[3]{##3}\VANthebibliography}

\usepackage{graphicx}	
\usepackage{amsmath}	
\usepackage{amssymb}	

\title[Radio Background Power Spectrum]{Measurement of the Anisotropy Power Spectrum of the Radio Synchrotron Background}

\author[A.R.~Offringa et al.]{A.R. Offringa$^{1,2}$\thanks{E-mail: \texttt{offringa@astron.nl}}, J.~Singal$^{3}$, S.~Heston$^{4}$, S.~Horiuchi$^{4,5}$, D.M.~Lucero$^{4}$\\ 
$^{1}$Netherlands Institute for Radio Astronomy (ASTRON), 
Oude Hoogeveensedijk 4, 7991 PD Dwingeloo, Netherlands\\
$^{2}$Kapteyn Astronomical Institute, 
P.O. Box 800, 9700 AV Groningen, Netherlands\\
$^{3}$Physics Department, University of Richmond, 
138 UR Drive, Richmond, VA 23173, USA\\
$^{4}$Department of Physics, Virginia Tech University, Blacksburg, VA 24061-0435, USA \\
$^{5}$Kavli IPMU (WPI), UTIAS, The University of Tokyo, Kashiwa, Chiba 277-8583, Japan}

\date{Accepted XXX. Received YYY; in original form ZZZ}

\pubyear{2021}

\begin{document}
\label{firstpage}
\pagerange{\pageref{firstpage}--\pageref{lastpage}}
\maketitle

\begin{abstract}
We present the first targeted measurement of the power spectrum of anisotropies of the radio synchrotron background, at 140~MHz where it is the overwhelmingly dominant photon background.  This measurement is important for understanding the background level of radio sky brightness, which is dominated by steep-spectrum synchrotron radiation at frequencies below $\nu \sim 0.5 \mathrm{~GHz}$ and has been measured to be significantly higher than that which can be produced by known classes of extragalactic sources and most models of Galactic halo emission.  We determine the anisotropy power spectrum on scales ranging from 2$^{\circ}$ to 0.2\arcmin\ with LOFAR observations of two 18~deg$^2$ fields --- one centered on the Northern hemisphere coldest patch of radio sky where the Galactic contribution is smallest and one offset from that location by 15$^{\circ}$.  We find that the anisotropy power is higher than that attributable to the distribution of point sources above 100~$\mu$Jy in flux.  This level of radio anisotropy power indicates that if it results from point sources, those sources are likely at low fluxes and incredibly numerous, and likely clustered in a  specific manner. 
\end{abstract}




\section{Introduction}\label{intro}
\begin{figure}
\begin{center}
\includegraphics[width=0.47\textwidth]{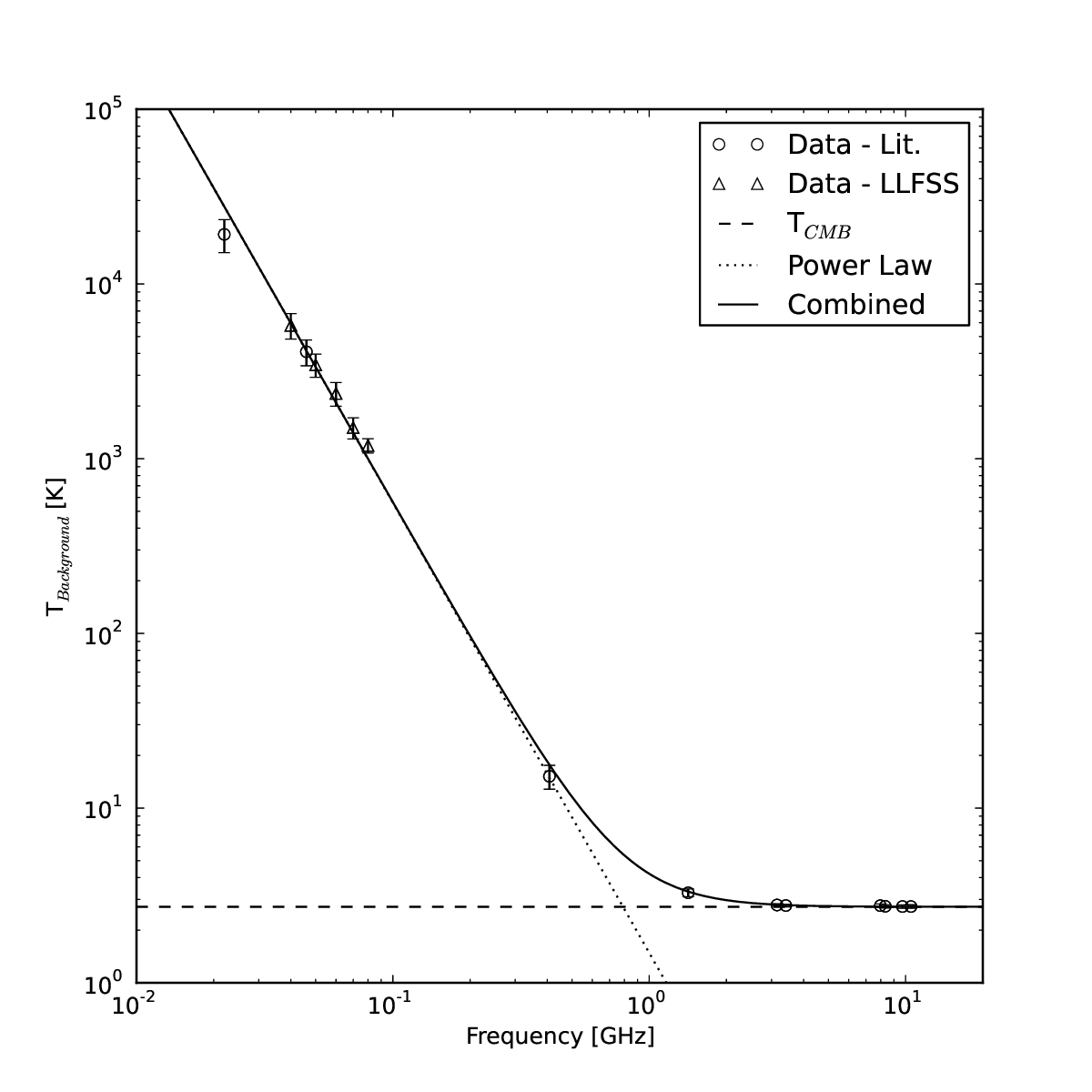} 
\end{center}
\vspace{-5mm}
\caption{The measured radio background brightness spectrum in radiometric temperature units reproduced from \citet{DT18}, as measured by the few measurements and maps where an absolute zero-level calibration was either explicit or obtained, including that work.  The brightness shows a clear power-law rise at frequencies below $\sim$10 GHz, above the otherwise dominant cosmic microwave background level represented by the dashed line.   } 
\label{f1}
\end{figure}

A puzzling question to have recently emerged is the origin of the radio background radiation.  The background level of radio sky brightness, which is due to some as of now unknown combination of integrated extragalactic sources and a possible large-scale Galactic halo, is dominated by steep-spectrum synchrotron radiation at frequencies below $\nu \sim 0.5 \mathrm{~GHz}$, and at higher frequencies it is present along with the otherwise dominant cosmic microwave background (CMB).  An apparent bright low-frequency background was reported as early as the 1960s \citep[e.g.,][]{Bridle67} and 1980s \citep[e.g.,][]{Haslam}.  Interest in this background was renewed by the surprisingly high absolute sky temperature at $\nu \sim 3 \mathrm{~GHz}$ reported by the ARCADE~2 \citep{Singal11} stratospheric balloon experiment.  Combining the ARCADE~2 measurements from 3--90$\mathrm{~GHz}$ \citep{Fixsen11} with several radio maps at lower frequencies from which an absolute zero-level has been inferred  \citep[recently summarized in][]{DT18} reveals a synchrotron background brightness spectrum
\begin{equation}
T_\mathrm{BGND}(\nu) = 30.4 \pm 2.6 \mathrm{K} \, \left(
\frac{\nu}{310\mathrm{~MHz}}
\right)^{-2.66 \pm 0.04} \, + \, T_{\rm CMB}
\label{T_B}
\end{equation}
shown in Figure~\ref{f1}, where $T_{\rm CMB}$ is the frequency-independent contribution of 2.725~K due to the CMB.  Following recent works we refer to this as the radio synchrotron background (RSB).

The reported bright background level is now in extreme tension with estimates of its expected level from the known radio emission mechanisms in the Universe, as recently summarized in \citet{CP}.  Several works have considered deep radio source counts and limited the integrated surface brightness from known classes of extragalactic radio sources to only around one-fifth of the radio background brightness level \citep[e.g.,][]{Vernstrom14,Condon12} including recently at 144~MHz \citep{Hardcastle20}.  Thus to achieve the measured radio background level from point sources would require an entirely new, incredibly numerous, heretofore unobserved population of low-flux radio sources.  As an alternative, various types of diffuse extragalactic sources such as cluster mergers \citep[e.g.,][]{FL15} and intergalactic dark matter decays and annihilations in galaxies, clusters, and filaments \citep[e.g.,][]{Fornengo11,Hooper14} have been proposed. 
Alternatively, a large, bright, roughly spherical synchrotron halo surrounding our Galaxy could explain part of the background \citep[e.g.,][]{SC13}. However such a large, bright halo would make our Galaxy unique among nearby spiral galaxies \citep{RB1} and would overturn our current understanding of the high-latitude Galactic magnetic field \citep{Singal10}.

One realm in which the RSB is almost completely unexplored is in its anisotropy.  Studies of temperature anisotropy power spectra have helped confirm the source populations responsible for the cosmic infrared \citep[e.g.,][]{Planck11,George15}  and gamma-ray \citep[e.g.,][]{Broderick14} backgrounds, and have been the most important component of CMB science thus far \citep[e.g.,][]{WMAP13}.  

The only direct constraints available in the literature on the anisotropy of the RSB at the most relevant angular scales are from confusion noise limits at a few discrete scales and based on measurements where considerations of the radio background in particular were incidental.  These include decades-old measurements in the GHz range where it is overwhelmed by the CMB by an order of magnitude or greater; specifically, based on observations made with the VLA at 8.4\,GHz \citep{Partridge97} and 4.9\,GHz \citep{Fomalont88}, and the Australia Compact Telescope Array at 8.7\,GHz \citep{Sub00}.  There are also recent measurements of the sky power spectrum at 150 MHz in several fields made with the Giant Metrewave Radio Telescope recently presented in \citet{Choudhuri20}.  The results presented there are in a more limited angular scale range, and in fields with higher Galactic diffuse emission structure contribution, than those presented here, and did not directly address specifically the question of the anisotropy power of the RSB.  At larger angular scales than those considered here, where the angular power is dominated by the large scale Galactic diffuse synchrotron structure, there are determinations of the angular power at 408~MHz reported by \citet{LaPorta08}. 

In this work we present a power spectrum of measured anisotropies of the RSB over the angular range from 2$^{\circ}$ to 0.2\arcmin\ based on dedicated LOw Frequency ARray \citep[LOFAR--][]{VH13} observations at 140~MHz of two 18~deg$^2$ fields. \S \ref{obs} describes the observations and data reduction and analysis methods, \S \ref{ps} presents the resulting power spectra, \S \ref{pops} explores possible point source populations that could produce the measured anisotropy power, and \S \ref{disc} presents a discussion.  Appendix \ref{appsec} provides a reference for considering the conversion factors between different computations and scalings of angular power that are relevant when bridging regimes and methods of determination where different conventions are in use.

\section{Observations and Data Reduction} \label{obs}

\begin{figure}
\begin{center}
\includegraphics[width=0.45\textwidth]{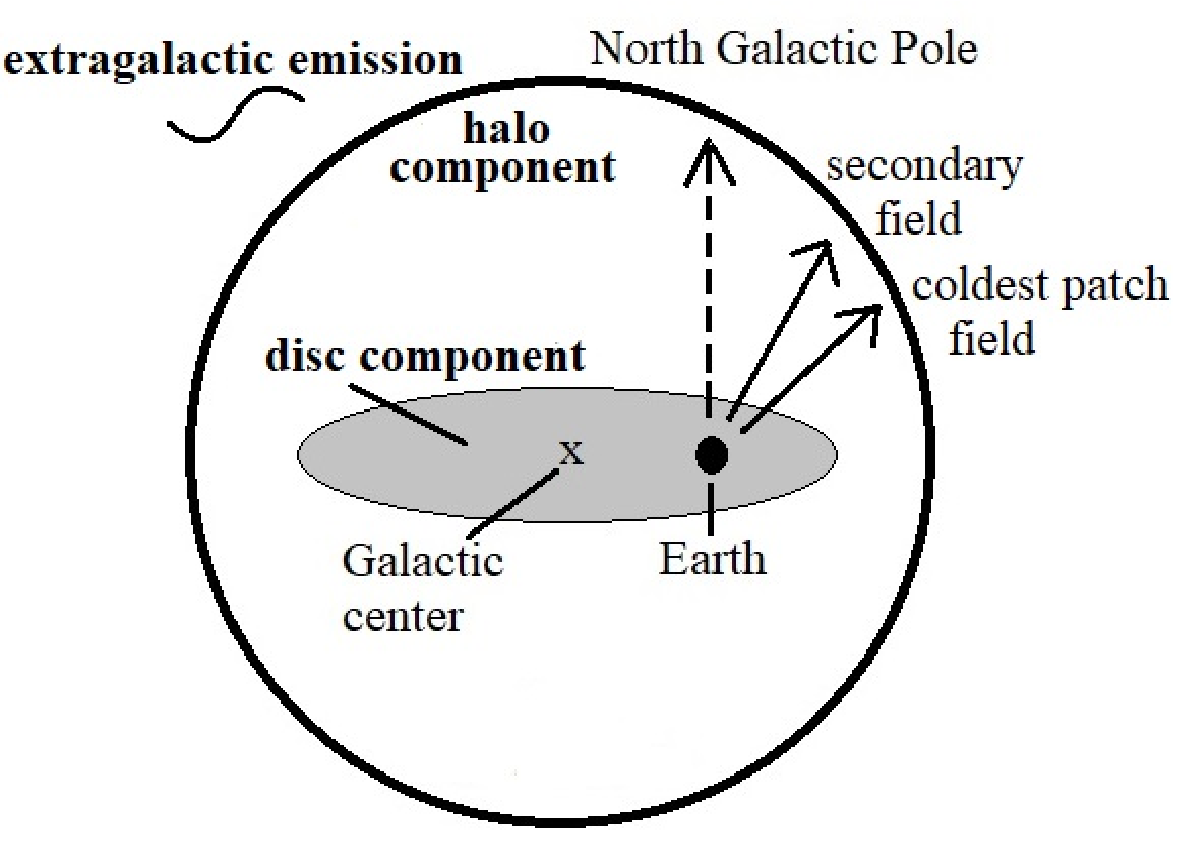}
\end{center}
\vspace{-5mm}
\caption{Schematic representation of the coldest patch and secondary target fields observed in this work in the context of a simple model of Galactic diffuse radio emission consisting of an ellipsoidal plane-parallel component due to the Galactic disc and a larger, spherical halo component, each centered on the Galactic center.  The coldest patch target field is in the direction of minimal integrated line-of-sight total contribution from the two components in the Northern Galactic hemisphere.  Such two-component models of large-scale diffuse Galactic radio emission are commonly utilized \citep[e.g.,][]{SC13,RB1}. } 
\label{lines}
\end{figure}

We use data from eight hours of dedicated observing with LOFAR in high band antenna (HBA) dual mode with Dutch stations only (23 core, 14 remote) in the band from 110-190~MHz on November 27, 2019.  As optimally this measurement should be done on a region with the minimum amount of Galactic diffuse foreground spatial structure, we chose a field centered on the Galactic Northern Hemisphere ``coldest patch'' (\citealt{Kogut11}; $9h\,38m\,41s$ +$30^{\circ}49'12''$, $l=$196.0$^{\circ}$ $b$=48.0$^{\circ}$), the region of lowest measured diffuse emission absolute temperature and thus where the integrated line-of-sight contribution through the Galactic components is minimal.  LOFAR allows simultaneous observation of an additional field offset by 15$^{\circ}$ in an adjacent 48~MHz wide band, so we chose a location toward the North Galactic Pole from the coldest patch of $10h\,25m\,00s$ +$30^{\circ}00'00''$ ($l$=199.0$^{\circ}$ $b$=57.9$^{\circ}$) which should have a slightly higher but still nearly minimal total Galactic contribution.  Fig.~\ref{lines} shows a schematic representation of the observed fields relative to a commonly employed simple model of the Galactic diffuse radio emission structure. The data cube consists of 666~baselines (all pairs of correlations) with four linear polarization pairs per visibility in 243~frequency channels with two-second integrations for a total of 1.4~TB of data per target field.  The 243~frequency channels are of equal width of 180~KHz and the filtering is done with a polyphase filter bank.  In addition to the target fields we observed the flux calibrator 3C\,295.

\begin{figure}
\begin{center}
\includegraphics[width=0.47\textwidth]{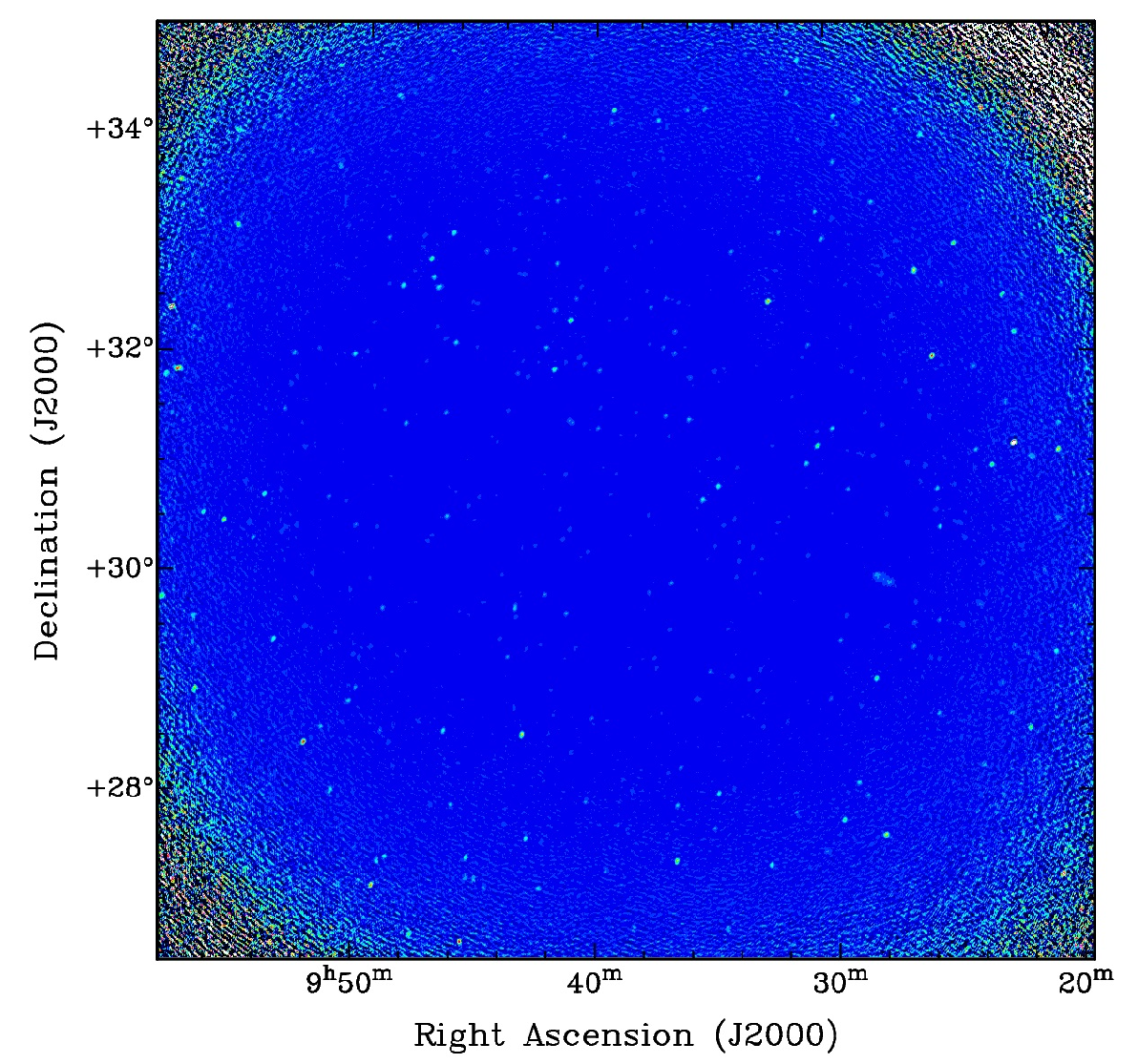}
\end{center}
\caption{An image of the coldest patch target field resulting from the imaging procedure discussed in \S \ref{obs}.  The synthesized beam measures 1.5\arcmin $\times$ 1\arcmin.  This field contains 3.6~Jy source 4C\,32.30 which can be used for self-flux calibration and an extended FRII galaxy (visible just to the lower right of the middle of the field).  All sources are removed for power spectrum determination, as point sources manifest power on all angular scales. } 
\vspace{-5mm}
\label{cop}
\end{figure}

Because we are interested in scales that are much larger than the effect of ionospheric activity, we only perform direction-independent calibration, thereby avoiding the effect of signal suppression that might incur during direction-dependent calibration. We have used two different methods for direction-independent calibration and imaging. The first approach is to use  Prefactor\footnote{\url{https://github.com/lofar-astron/prefactor}}, the standard automated LOFAR direction-independent calibration pipeline \citep{Weeren_2016,williams_2016}, which makes use of several software packages including the Default Pre-Processing Pipeline (DP3; \citealt{dp3-van-diepen-2018}), LOFAR SolutionTool (LoSoTo; \citealt{losoto-2019}) and AOFlagger \citep{offringa-2012-scale-invariant-rank-operator}. Because this pipeline has not been developed for power-spectrum experiments, for verification we also calibrate our data in a manual approach.

Manual calibration is only performed for the coldest patch field. For manual calibration, we start with running \textsc{aoflagger} to flag outlying data points due to RFI contamination and one outlying station.  We perform initial flux and phase calibrations for each sub-band using the flux calibrator observation which are then applied to the target fields.  We image the target fields with \textsc{wsclean} \citep{AO14} with primary beam correction and standard \textsc{clean} settings to extract an initial point source model for self-calibration.  Specifically, we first run the source extractor \textsc{aegean} \citep{Hancock18} with a high flux threshold of 9$\sigma$ to extract a shallow model, containing about 300 sources.  Following this, we run self-calibration using the model on 25 sub-bands and image at a higher resolution of 45\arcsec.  We next re-run Aegean with a lower flux threshold (7$\sigma$), extracting 2396 sources.  

Many of these sources are near the edge of the primary beam and so are likely false detections. We cut out all sources that are at a place in the image where the beam has less than 5\% gain, reducing the model to 644 sources with almost no false positives.  Extended sources, including one prominent FRII galaxy, are excluded from calibration.  Source 4C\,32.30 with flux 3.6~Jy \citep{Waldram96} is near the middle of the coldest patch target field and is used for flux calibration. Using these full calibration solutions we re-image the target field with \textsc{wsclean} with 20\arcsec\ resolution. An image of the coldest patch target field is shown in Fig.~\ref{cop}. 

The results of the automated and manual approach are found to be similar on the coldest patch field, so we process the secondary target field only with the automated approach using Prefactor. Before making power spectra, we subtract the foreground sources using a deep \textsc{wsclean} multi-frequency deconvolution using auto-masking \citep{offringa-2017}.  The auto-masking ensures that all sources $\ge$7$\sigma$ are subtracted to a 1$\sigma$ level. To avoid subtracting a diffuse component, we do not use multi-scale clean.  Removing sources $\ge$7~$\sigma$ with RMS noise of 720~$\mu$Jy results in sources above 5~mJy being removed.  

\section{Power Spectrum} \label{ps}

\subsection{Full angular power spectrum}

Our angular power spectrum pipeline is based on the pipeline described by \citet{AO19}, which is originally written for the LOFAR Epoch of Reionization project \citep{mertens-2020}. Our angular pipeline produces a power spectrum of fluctuations from the source-removed target field images.  We use the central 4.3$^{\circ}$
square of each target field image for determination of the angular power spectra.  A quantitative discussion of the procedure for forming a power spectrum from an interferometric image is presented in Appendix \ref{appsec}.  The steps in making a power spectrum are:
\begin{itemize}
\item[--] Make a naturally weighted image using multi-frequency synthesis from the source-subtracted data. We use \textsc{wsclean} for this with increased accuracy settings (see \citealt{offringa-2019}).
\item[--] Convert the flux density image to units of temperature (Kelvin) using Eq.~(\ref{eq:jansky-to-kelvin}).
\item[--] Take the spatial Fourier transform of the image and point spread function (PSF) to create a complex ($u$,$v$) grid for both.
\item[--] Elementwise divide the complex $uv$ image by the complex value of the $uv$ PSF.
\item[--] Average the power in annuli and normalize these.
\end{itemize}
This method of determining the power spectrum, where we correct for the (in our case, natural) image weighting function in $uv$-space, alleviates the need to perform a bias-correction of the power spectrum.  Otherwise, image-based reconstruction of power spectra can give a biased estimate of the true sky signal due to the correlated noise in the image domain \citep{DM19}.

\begin{figure}
\begin{center}
\includegraphics[width=0.47\textwidth]{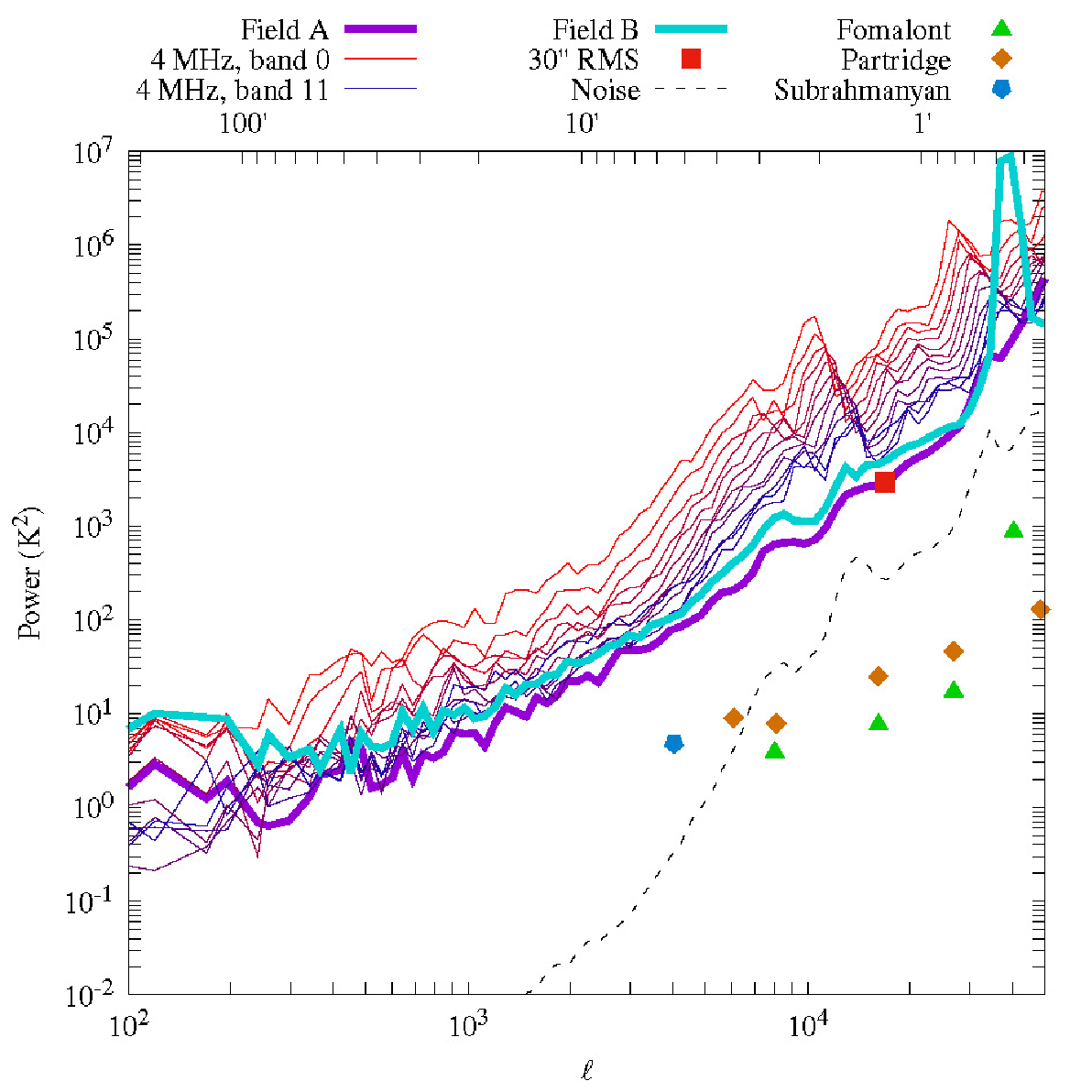}
\end{center}
\vspace{-2mm}
\caption{Measured anisotropy power spectrum of the radio sky centered at 140 MHz with 720 $\mu$Jy RMS noise.  Shown are curves for the full bandwidth of the coldest patch field (field A) and the secondary field (field B), as well as for 12 four MHz wide sub-bands of field A.  The anisotropy in field A deduced by considering the average noise per beam in the image with the synthesized beam tapered to 30\arcsec\ FWHM is also shown and agrees at the relevant angular scale, given by Eq. (\ref{fwhmeq}).  We also show comparison levels inferred by the noise per beam at 8.7~GHz, 8.4~GHz, and 4.9~GHz in different fields as calculated by \citet{Holder14} and scaled here to 140~MHz assuming a synchrotron power law of -2.6 in radiometric temperature units.  The amount of angular power is $\sim$1.4 times higher for field B compared to field A (in K$^2$ units) across a range of angular scales, as discussed in \S \ref{disc}.  All angular powers are expressed here in the $\left({\Delta T} \right)_\ell^2$ normalization.
} 
\label{meas-specs}
\end{figure}

The resulting power spectrum of fluctuations is shown in Fig.~\ref{meas-specs}, for both the coldest patch and secondary target fields.  We also show the spectrum of fluctuations calculated for 12 four MHz wide sub-bands separately.  The lowest frequency sub-band has approximately 17 times more power in these K$^2$ units than the highest frequency sub-band because of the spectral dependence of synchrotron radiation (c.f., Eq.~\ref{T_B}), which is a bit over 4 times as bright in radiometric temperature units (and thus around 17 times as bright in K$^2$ units) at 190~MHz than at 110~MHz.  The angular power is lower in the full bandwidth because of more complete $u-v$ coverage. 

These measurements will have a contribution from the noise of the instrument. To calculate this contribution, we extract Stokes-V visibilities and measure the differential variance between consecutive Stokes-V visibilities in time, multiply by $2$ and assume that this is a representative noise value for all visibilities. We calculate the noise power spectrum by replacing all visibilities by randomly sampled Gaussian values with the calculated variance, produce images and calculate a power spectrum from the images as described. The result is shown as dashed line in Fig.~\ref{meas-specs}. Given that the noise power is at least an order of magnitude below the measured power, clearly our measured power is dominated by something other than the system noise.  Additionally, while one would expect the power in the full band to be the average of the sub-bands for the case of noise-dominated power, in this case with the noise contribution sub-dominant the more complete u-v coverage of the full band will result in lower angular power.

\subsection{Power from RMS fluctuations}

We can also calculate the  power on a specific, discrete angular scale in a completely different, complementary way, following a procedure discussed in \citet{Holder14}:  

\begin{itemize}
\item[--] The noise per beam in the image $\Delta S_\textrm{Jy/psf}$ is measured with the synthesized beam tapered to 30\arcsec~full with at half maximum (FWHM). The noise level of this image is 720 $\mu$Jy.
\item[--] The beam is fitted to an elliptical Gaussian with major and minor axes $w_{\rm maj}$ and $w_{\rm min}$ to calculate the synthesized beam solid angle in radians,
\begin{equation}
    \Omega_\textrm{psf}=\pi (w_{\rm maj} \times w_{\rm min}) \times \left(
    \frac{1}{60}
    \right)^2 \times \left(
    \frac{\pi}{180}\right)^2 \times \left(\frac{1}{4 \log 2}\right).
\label{omegapsf}
\end{equation}    
\item[--] The resulting temperature fluctuation $\Delta T$ is calculated by
\begin{equation}
 \Delta T = \Delta S_\textrm{Jy/psf} \frac{ 10^{-26} c^2 }{ 2 k_B \nu^2 \Omega_\textrm{psf}},
\label{Tpsf}
\end{equation} 
to achieve ${{\Delta T}}$ on a the angular scale corresponding to a Gaussian beam of 30\arcsec~FWHM.  
\end{itemize}
The fluctuation power calculated in this way is shown in Fig.~\ref{meas-specs} for the coldest patch target field.  The angular scale corresponding to a 30\arcsec~FWHM Gaussian beam does not exactly match a 30\arcsec~spherical harmonic due to the beam taper, which is why the RMS measurement is not exactly placed at 30\arcsec~(see upper horizontal axis in Fig.~\ref{meas-specs}). To calculate the corresponding angular scale we have used the formula provided in \citet{Holder14} which gives the $\ell$ value corresponding to a particular synthesized beam FWHM expressed in radians:
\begin{equation}
    \ell = \frac{2.35} {\rm FWHM}.
    \label{fwhmeq}
\end{equation}
We see that the fluctuation power calculated directly from the noise per beam matches that determined from a full power spectrum at the particular angular scale.  We also show the fluctuation power obtained in this way by \citet{Holder14} from the measurements of \citet{Partridge97},  \citet{Fomalont88}, and \citet{Sub00}, scaling from the relevant GHz frequencies to 140~MHz by a synchrotron power law of -2.6 in radiometric temperature units. 

\subsection{Power from potentially unremoved point sources}
\label{unrem}

In order to quantify the contribution of potentially unremoved point sources in the images to the measured angular power, we created a Monte Carlo catalog of simulated sources. We interpolate the simulated sources onto a grid (using sinc-interpolation) and simulate visibilities from the resulting sky image to apply the instrumental effects that affect the power spectrum ($uv$-sampling and the primary beam). We use the Image Domain Gridder (IDG; \citealt{vandertol-idg-2018}) inside \textsc{wsclean} to apply the time and frequency dependent LOFAR primary beam. The resulting visibilities are processed with our imaging and power spectrum generation pipeline. 

\begin{figure}
\begin{center}
\includegraphics[width=0.47\textwidth]{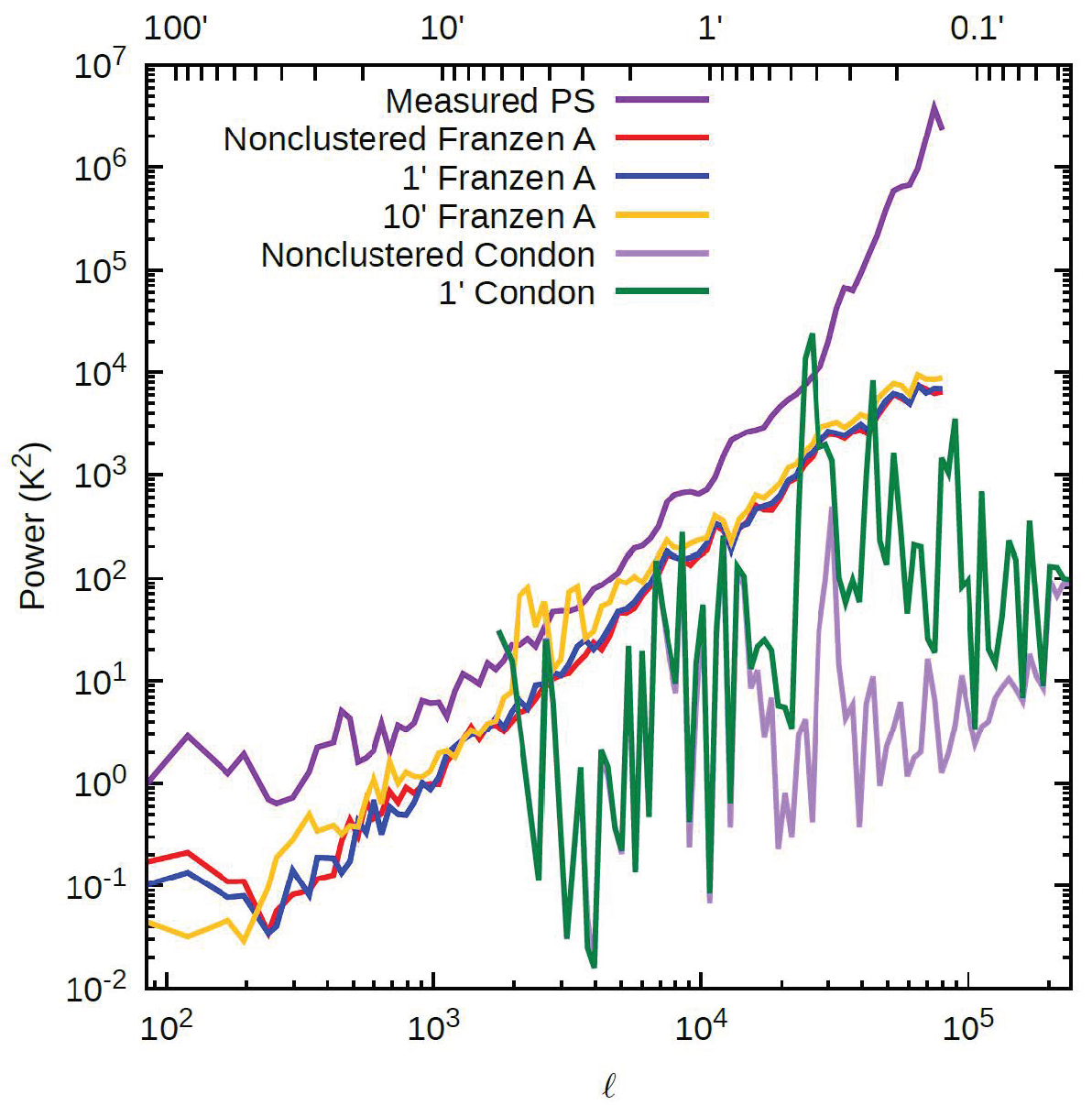}
\end{center}
\caption{Measured power spectrum (for the coldest patch target field) and the simulated full-pipeline anisotropy power spectrum resulting from i)  potential unremoved point sources down to 100~$\mu$Jy according to a point source model presented in \citet{Franzen16} discussed in \S \ref{unrem}, ii) the same \citet{Franzen16} model with sinusoidal clustering added on scales of 1\arcmin\ and 10\arcmin, iii) potential point sources down to nano-Jansky fluxes according to the point source model presented in \citet{Condon12} discussed in \S \ref{pops} which can reproduce the surface brightness level of the RSB, and iv) the same \citet{Condon12} model with sinusoidal clustering added on a scale of 1\arcmin. } 
\label{sim-specs}
\end{figure}

We distributed the sources in flux (S) between 100 $\mu$Jy and 10~mJy  according to four models presented in \citet{Franzen16}, based on their measured and extrapolated deep source counts at 150~MHz, of the form
\begin{eqnarray}
    n(S)=\frac{dn}{dS}=k_1 \left(\frac{S}{Jy}\right)^{\gamma_1} {\rm Jy^{-1} \, Sr^{-1} \, for \,\, 0.1 \, mJy < S < 6.0 \, mJy} 
\\
\nonumber    = k_2 \left(\frac{S}{Jy}\right)^{\gamma_2} {\rm Jy^{-1} \,  Sr^{-1} \, for \,\, 6.0 \, mJy < S < \, 10 \, mJy}
    \label{modeldist}
\end{eqnarray}
and randomly in RA and Dec, with frequency spectral indexes distributed normally around -2.6 in radiometric temperature units with a standard deviation of 0.1.  The parameters $k_1, k_2, \gamma_1$ and $\gamma_2$ are given for four models in Table 2 of that work for the extrapolated portion.  The four models, which are identical in the number of high flux sources but differ in the number of low flux sources, result in roughly the same amount of angular power at all scales, indicating that the contribution to the angular power is dominated by the higher flux sources in this flux range.  We then introduce a simple sinusoidal spatial clustering in both RA and Dec on scales of 1\arcmin\ and 10\arcmin\ to see whether clustering can have a significant effect on the observed angular power resulting from this model, with results for one of the models (model ``A'') visualized in Figure~\ref{sim-specs}. 

The clustering on a 1\arcmin\ scale has very little effect on the measured angular power on any angular scale for this model or any of the four.  This is because 1\arcmin\ is considerably less than the average separation of the higher flux sources which primarily contribute to the measured angular power.   In fact the clustering added in this way on the 1\arcmin\ scale slightly reduces the angular power on some angular scales because the angular power involves circular averaging while the sinusoidal variation has been added to essentially rectangular RA and Dec coordinates at this scale.  The 10\arcmin\ clustering manifests an appreciable increase in angular power on that particular scale.  However this only propagates to some smaller angular scales, and we see from the 1\arcmin\ clustering case that below a certain angular scale (somewhere between 10\arcmin\ and 1\arcmin), clustering in these models cannot add further angular power.

Given that the contribution to the angular power is dominated by the higher flux sources in these models, this modeled observed power resulting from unremoved sources above 100~$\mu$Jy indicates that unsubtracted point sources in the images above the flux detection limit are not a major contributor to the measured angular power, and that sources above 100~$\mu$Jy generally cannot produce the measured angular power on at least some angular scales.

\section{Possible Source Population} \label{pops}

With the angular anisotropy power of the radio sky being  larger than that which can be accounted for by point sources above 100~$\mu$Jy, the question arises as to what flux source count distributions (often denoted $n({\rm S})$ or $\frac{dn}{d{\rm S}}$) of faint point sources could give rise to the measured angular power.  As discussed in \citet{Condon12}, regarding the surface brightness of the RSB, if it is indeed that given by Eq.~(\ref{T_B}), then if originating from point sources, given the measured constraints on the source counts above 10~$\mu$Jy, those sources must be lower flux and incredibly numerous.  We will consider here the possibility that the angular power as well is due to a large number of low-flux point sources.

\citet{Condon12} present three hypothetical low-flux point source population flux distributions which could provide the measured surface brightness of the RSB.  These distributions are of the approximate form
\begin{equation}
    n({\rm S})=\frac{A}{\rm S^2}\exp{\left(-4 \ln (2) \frac{\left[\log ({\rm S}) - \log ({\rm S_{pk}})\right]^2}{\phi^2}
    \right)} \, \,  {\rm Jy^{-1} \, Sr^{-1} }
    \label{Cdisteq}
\end{equation}
with the normalization $A$, width $\phi$, and the flux of the peak contribution to the background per log flux bin ${\rm S_{pk}}$ given for the three models.  All three models feature a large number of low flux sources with values of ${\rm S_{pk}}$ of approximately 0.03 $\mu$Jy, 0.02 $\mu$Jy, and 0.003 $\mu$Jy and a density of sources on the sky exceeding that measured in the Hubble Ultra Deep Field by at least an order of magnitude.  

In a procedure similar to the simulation discussed in \S \ref{unrem} we simulate point source populations distributed in flux according to Eq.~(\ref{Cdisteq}) with frequency spectral indexes distributed normally around -2.6 in radiometric temperature units with a standard deviation of 0.1 and run the resulting simulated sky through our simulation, imaging, and power spectrum generation pipeline as described there.  We adopt the model with the smallest number of sources, which still results in an average  around 200 million sources in 1$^{\circ}$ square, or 150 sources per pixel.  Due to the very large number of sources in this model, computing limitations require the simulated field of view to be smaller, 0.5$^{\circ}$ on a side, so calculation of angular power on angular scales larger than this is not possible.  The resulting power can still be calculated for most of the range of angular scales of relevance depicted in e.g., Fig.~\ref{meas-specs}.  The result for sources distributed randomly in RA and Dec is shown in Fig.~\ref{sim-specs}.  It is seen that because of the very large average number of sources per pixel the proportional variation in brightness among pixels is small and therefore the resulting angular power in this isotropic model is low.  To preliminarily investigate the effects of clustering for this model we adopt the same simple sinusoidal clustering on a 1\arcmin scale as  discussed in \S \ref{unrem}, with results also shown in Fig.~\ref{sim-specs}.  In this case, with the very large number of sources, the added clustering increases the simulated observed angular power significantly on all angular scales that are equal to and smaller than that of the clustering scale (corresponding in this case to $\ell \sim $ 11000). Thus, we conclude that it is a possibility that with the appropriate clustering on many angular scales over a wide range, the \citet{Condon12} model of many very low flux sources could reproduce the observed angular power.

\section{Discussion} \label{disc}
 
We have carried out a dedicated measurement with LOFAR to determine the anisotropy angular power of the radio background at 140~MHz on angular scales ranging from 2$^{\circ}$ to 0.2\arcmin.  As discussed in \S \ref{obs} our results stem from eight hours of observing of two fields with a minimal amount of Galactic diffuse foreground structure.  As shown in \S \ref{ps} both the direct method of imaging, removing sources, and calculating the power spectrum, and the method of considering the noise per beam in the image with the synthesized beam tapered to a specific width, yield a measured angular power that is more than that which would result from point sources above 100~$\mu$Jy, either distributed randomly spatially or clustered.  As shown in Fig.~\ref{meas-specs} the angular power is also at least an order of magnitude larger than that inferred from measurements at GHz frequencies.  

Our measured angular power is around a factor of $\sim$3 (in the $\Delta T$ normalization) {\it lower} than that reported by \citet{Choudhuri20} in the angular scales of overlap (from $10^2 \leq \ell \leq 3 \times 10^3$), applying Eq.~(\ref{T2}) to convert to the plotted $C_{\ell}$ units of their Fig.~1.  They observe four fields at a variety of Galactic latitudes and longitudes, with all looking through significantly more Galactic structure than the fields in this work.  Their reported angular power differs somewhat at various reported $\ell$ values over their different fields, but this manifests no apparent correlation with the amount of Galactic structure along a line of sight, indicating that the discrepancy between their fields, and more relevantly with the results here, may be due to instrumental effects and analysis considerations.  Interestingly, our measured angular power quite closely matches the {\it modeled} angular power of unsubtracted point sources below 50~mJy reported in \citet{Choudhuri20}.

The angular power measured here is due to a combination of that due to extragalactic sources, that due to structure in Galactic diffuse emission, and, in principle, possible considerations such as RFI, sidelobe pickup, and residual power from subtracted sources stemming from calibration errors.  The lack of artifacts in the visibilities indicate that RFI is not a significant contributor to this measurement, and we don't see evidence of sidelobe pickup in the images, as the sidelobe positions are frequency-dependent and would thus present as shifting patterns in each sub-band.  It is the case that with currently available LOFAR analysis techniques we cannot absolutely rule out a contribution from the residual power from subtracted sources stemming from calibration errors.  In particular, making power spectra at larger scales presents a particular calibration challenge in this regard, as also found by epoch of reionization measurements \citep{SK18,Patil16,Barry16}.  Future development of LOFAR analysis techniques may allow a more precise determination of this, but these are beyond the scope of the present work.

We can estimate the contribution due to structure in Galactic diffuse emission by noting that the measured angular power in the secondary field is systematically a factor of $\sim$1.4 higher in the $(\Delta T)^2$ normalization than that in the coldest patch field, as seen in Fig.~\ref{meas-specs}, and thus a factor of $\sim$1.2 higher in the $(\Delta T)$ normalization.  This is, quite tellingly, the same as the square of ratio of the average absolute brightness in radiometric temperature (K) units for the two regions that we calculate using the \citet{Haslam} map averaging over pixels within 4$^{\circ}$ from the field centers ($1.2 \pm 0.1$).  As the differences in absolute brightness are due solely to differences in lines of sight through the Galactic diffuse components (as visualized in Fig.~\ref{lines}), this is a strong indication that the proportion of angular power in $(\Delta T)$ units due to Galactic structure tracks the proportion of absolute brightness due to that structure, for lines of sight in this general direction of minimal Galactic structure and likely for general lines of sight far away from the Galactic plane.  A number of considerations point to the extragalactic component being overwhelmingly dominant (by at least a factor of 5) in terms of the absolute temperature of the background \citep[e.g.,][]{RB1} so we believe that the extragalactic component dominates the measured angular power in the coldest patch field by approximately this factor.  Stated another way, the {\it normalized} angular power $\left(\frac{\Delta T}{T}\right)$ for both fields is the same, indicating that the contribution to the angular power from Galactic structure is sub-dominant when considering these fields since it is the component that varies spatially between the two fields.  

If the angular power measured here is due to low flux radio point sources, they must be very numerous, paralleling the situation when considering the surface brightness of the radio background.  As discussed in \S \ref{pops} we simulated the angular power resulting from a source count distribution representing a very large number of sources below 1~$\mu$Jy that, as shown in \citet{Condon12}, could potentially provide the level of surface brightness of the RSB.  We found that this source distribution when distributed randomly spatially contains low angular power due to the large number of sources per pixel, but could possibly reproduce the angular power of the RSB measured here given the proper detailed clustering on a wide range of angular scales.  It is our intention to continue this modeling in a future work in order to determine the precise clustering parameters for very large numbers of low-flux sources which could, possibly, result in the angular power spectrum observed here, and to explore the implications of such a population.

\section*{Acknowledgements}

We thank the ``BAM - Anisotropic Universe 2018'' workshop where initial discussions took place. We acknowledge LOFAR award LC12-005. S.~Heston is supported NSF Grant No.~PHY-1914409. S.~Horiuchi is supported by the U.S.~Department of Energy Office of Science under award number DE-SC0020262 and NSF Grants No.~AST-1908960 and No.~PHY-1914409. This work was supported by World Premier International Research Center Initiative (WPI), MEXT, Japan. This paper is based on data obtained with the International LOFAR Telescope (ILT) under project code LC12-005. LOFAR \citep{VH13} is the Low Frequency Array designed and constructed by ASTRON. It has observing, data processing, and data storage facilities in several countries, that are owned by various parties (each with their own funding sources), and that are collectively operated by the ILT foundation under a joint scientific policy. The ILT resources have benefited from the following recent major funding sources: CNRS-INSU, Observatoire de Paris and Université d'Orléans, France; BMBF, MIWF-NRW, MPG, Germany; Science Foundation Ireland (SFI), Department of Business, Enterprise and Innovation (DBEI), Ireland; NWO, The Netherlands; The Science and Technology Facilities Council, UK.

\section*{Data Availability}

The data underlying this article will be shared on reasonable request to the corresponding author.


\bibliographystyle{mnras}




\appendix

\section{Relations Between Measures of Angular Power of Temperature Anisotropies}\label{appsec}

In this appendix we present the scaling relationships between two normalizations for the angular power of temperature anisotropies, and derive which arises naturally from power spectra obtained from interferometric observations. 

\subsection{Power Spectrum Normalizations and Multipole Moments}
{\bf \noindent Note on notation and dimensions:}  Square brackets [ ] will indicate ``dimensions of.''  Here, in order to keep track of factors of the angular scale $\ell$ and normalization factors of $\pi$ in quantities, we will follow factors of $\ell$ and $\pi$ by a quasi-dimensionality.  That is, to encompass both dimensions with physical units and these normalization factors we will refer in this work to ``quasi-dimensionality'' to encapsulate both.  Definitions of relevant quantities have been obtained from \citet{Ryden}, \citet{Les}, and \citet{Jackson}.

The temperature fluctuation, or deviation from the average temperature, at a point on the sky in the direction $\hat{n}$ or equivalently at angular coordinates $(\theta,\phi)$, denoted ${{\delta T}}$, can be expressed with the spherical harmonic functions and their coefficients:
\begin{equation}
{{\delta T}}(\theta,\phi) \equiv {{T(\theta,\phi)-\langle T \rangle} } = \sum_{\ell,m} a_{l,m} Y_{l,m}(\theta,\phi) .
\label{e1}
\end{equation}
The `{\it angular correlation function}' $C(\theta)$ is an average of the product of $ {{\delta T}}(\hat{n})$ values in directions separated by the angle $\theta$:
\begin{equation}
C(\theta) = \left\langle{ {{\delta T}} (\hat{n}) \, \, {{\delta T}} (\hat{n}')  }\right\rangle_{\hat{n} \cdot \hat{n}'=\cos(\theta)}. 
\end{equation}
Therefore, quasi-dimensionally,
\begin{equation}
[ C(\theta) ] = \left [{{\delta T}} \right ]^2 \, .
\end{equation}
$C(\theta)$ can be expressed as a sum of Legendre Polynomials and `{\it multipole moment}' coefficients $C_{\ell}$:
\begin{equation}
C(\theta) = \frac{1}{4\pi} \sum_{\ell} (2\ell+1) \, C_{\ell} \, P_{\ell}(\cos \theta) .
\label{Ctheta}
\end{equation}
To assess the quasi-dimensionality of the Legendre Polynomials, we can use the Spherical Harmonics addition theorem
\begin{equation}
P_{\ell}(\cos \theta) = {{4\pi} \over {2\ell +1}} \sum_m {Y_{\ell,m}}(\theta,\phi) {Y^*_{\ell,m}(\theta',\phi')}
\end{equation}
Now we must note that any sum over $m$ for a given $\ell$ runs from $-\ell$ to $+\ell$ and so has $2\ell+1$ terms, and therefore such sums have a quasi-dimensionality of [$2\ell+1]$.  Therefore, quasi-dimensionally,
\begin{equation}
[P_{\ell}(\cos \theta)] = [4\pi] \cdot [ {Y_{\ell,m}}(\theta,\phi) ]^2. 
\end{equation}
To evaluate the quasi-dimensionalty of the spherical harmonics ${Y_{\ell,m}}(\theta,\phi)$, we can note simply that $Y_{0,0}=\frac{1}{\sqrt{4\pi}}$ so that $[{Y_{\ell,m}}(\theta,\phi)]=\left [{{1}\over{4\pi}}\right ]^{1/2}$, so $[P_{\ell}(\cos \theta)] =[\,]$; the Legendre polynomials are quasi-dimensionless.
Returning to Eq.~(\ref{Ctheta}) then, 
\begin{equation}
[ C_{\ell} ] = \left [{{4\pi} \over {2\ell+1}} \right ] \cdot [ C(\theta) ] = \left [{{4\pi} \over {2\ell+1}} \right ] \cdot \left [{{\delta T}} \right ]^2  .
\label{Cdem}
\end{equation}
The multipole moments $C_{\ell}$ are the variance (mean of the squares) of the spherical harmonic coefficients:
\begin{equation}
C_{\ell} =  \langle{ \left\vert a_{\ell,m} \right\vert^2  }\rangle = { {1} \over {2\ell + 1}}\sum_m \left\vert a_{\ell,m} \right\vert^2 \, ,
\end{equation}
where again the sum over $m$ for a given $\ell$ has a quasi-dimensionality of [$2\ell+1]$.  Thus,
\begin{equation}
[C_{\ell}] = \left [a_{\ell,m}\right ]^2,
\end{equation}
and, utilizing Eq.~(\ref{Cdem}),
\begin{equation}
[a_{\ell,m}] = \left [ C_{\ell} \right ]^{1/2} = \left [{{4\pi} \over {2\ell+1}} \right ]^{1/2} \cdot \left [{{\delta T}} \right ]
\label{rt}
\end{equation}
We note that this does not imply that $\left [{{\delta T}} \right ]$ as expressed in Eq.~(\ref{e1}) is "quasi-dimensionless" with respect to factors of $\ell$ -- i.e., it does not imply that $\left [{{\delta T}} \right ]$ has no natural scaling with $\ell$ -- rather we are just tallying the conversion factors between the quasi-dimensionality of various quantities.

Eq.~\ref{rt} tells us that in order to express the temperature fluctuation power at a given angular scale $\ell$ in terms of the multipole moment, we would need
\begin{equation}
\left( {{\delta T}} \right)_{\ell} = \sqrt{ {{2\ell+1} \over  {4\pi} }} \, \sqrt{C_{\ell}}.
\label{T1}
\end{equation}
However, a standard normalization scheme used in the cosmic microwave background literature \citep[e.g.,][]{Planck11} is to express the temperature fluctuation angular power multiplied by different factors of the angular scale $\ell$, resulting in a measure of the fluctuation power, here denoted $({{\Delta T}})_{\ell}$, that would be constant across values of $\ell$ in the case of an invariant spectrum of Gaussian random fluctuations:
\begin{equation}
\left( {{\Delta T}} \right)_{\ell} \equiv \sqrt{ {{\ell(\ell+1)} \over {2\pi}} }  \, \sqrt{C_{\ell}}.
\label{T2}
\end{equation}
With Eqs.~(\ref{T1}) and (\ref{T2}) we see that the relation between the two normalizations of the temperature fluctuation power is
\begin{equation}
\left( {{\Delta T}} \right)_{\ell} = \sqrt{ {{2\ell(\ell+1)} \over {2\ell+1}} } \left( {{\delta T}} \right)_{\ell}.
\end{equation}
We note that the quantities $({{\delta T}})_{\ell}$ and $({{\Delta T}})_{\ell}$ can be scaled to be normalized by the average temperature so that they express a fractional deviation from it, and are then denoted $\left( {{\delta T} \over {T}} \right)_{\ell}$ and $\left( {{\Delta T} \over {T}} \right)_{\ell}$, respectively.  We will now show that the power spectra produced from interferometric observations most naturally have a normalization of $({{\Delta T} })_{\ell}^2$.

\subsection{Power Spectra from Interferometric Observations}

To determine that the power spectra determined from interferometric observations are in the $({{\Delta T} })_{\ell}^2$ normalization, we first consider that the power spectrum relates to $\tilde{T}$, the Fourier transform of the temperature field, as:
\begin{equation} \label{eq:ps}
 P(\mathbf{k}) \equiv A \left| \tilde{T}(\mathbf{k}) \right|^2
\end{equation}
With $A$ the physical area of the field (in Mpc$^2$). In this equation, the Fourier transform is defined with a ``1/N'' normalization:
\begin{equation} \label{eq:def-ft}
 \tilde{T}(2\pi\mathbf{k}) \equiv \frac{1}{N_x N_y} \sum_\mathbf{x} T(\mathbf{x}) e^{-i 2 \pi \mathbf{k} \cdot \mathbf{x}}
\end{equation}

When a ``number of image pixels'', $N_x N_y$, is used, this is defined as the \textit{effective} number of pixels. The definition for $N_x N_y$ is:
\begin{equation} \label{eq:defN}
 N_x N_y = \frac{\Omega_A}{\Omega_\textrm{psf}},
\end{equation}
with $\Omega_A$ the primary beam solid angle, and $\Omega_\textrm{psf}$ the synthesized beam solid angle. For completeness, the conversion from flux density per beam (Jansky/psf) to temperature (Kelvin):
\begin{equation} \label{eq:jansky-to-kelvin}
 T(\mathbf{x}) \equiv S_\textrm{Jy/psf}(\mathbf{x}) \frac{ 10^{-26} c^2 }{ 2 k_B \nu^2 \Omega_\textrm{psf}},
\end{equation}
with $S$ in Jansky/PSF and $T$ in Kelvin. If Eqs.~\eqref{eq:defN} and \eqref{eq:jansky-to-kelvin} are substituted in Eq.~\eqref{eq:def-ft}, then $\Omega_\textrm{psf}$ cancels out:
\begin{equation} \label{eq:ft-written-out}
 \tilde{T}(2 \pi \mathbf{k}) = \frac{ 10^{-26} c^2 }{2 k_B \nu^2 \Omega_A} \sum_x S(\mathbf{x}) e^{-i 2 \pi \mathbf{k} \cdot \mathbf{x}}.
\end{equation}
Therefore, bringing everything together and assuming as input a correctly normalized image in units of temperature $T(\mathbf{x})$:
\begin{equation} \label{eq:ps-full}
 P(2 \pi  \mathbf{k}) = A \left| \frac{1}{N_x N_y} \sum_\mathbf{x} T(\mathbf{x}) e^{-i 2 \pi \mathbf{k} \cdot \mathbf{x}} \right|^2
\end{equation}

The power spectrum is often expressed in so-called ``dimensionless units'' \citep[e.g.,][]{YAH} which, perhaps confusingly, ends up with physical units of squared temperature (e.g. K$^2$). The dimensionless power spectrum relates to the two-dimensional power spectrum as follows:
\begin{equation}
 \Delta^2(k) = P(k) \frac{k^2}{2 \pi}.
\end{equation}
Substituting Eq.~\eqref{eq:ps-full} into this equation,
\begin{equation}
 \Delta^2(2 \pi \mathbf{k}) = 2 \pi k^2 A \left| \frac{1}{N_x N_y} \sum_\mathbf{x} T(\mathbf{x}) e^{-i 2 \pi \mathbf{k} \cdot \mathbf{x}} \right|^2
\end{equation}
The units for the transverse distance are arbitrary in this equation, as they cancel out through $k^2 A$. If we chose $p$ to be the dimensionless counterpart of $k$ such that $A=1$, then
\begin{equation}
 \Delta^2(2 \pi \mathbf{p}) = 2 \pi |\mathbf{p}|^2 \left| \frac{1}{N_x N_y} \sum_\mathbf{x} T(\mathbf{x}) e^{-i 2 \pi \mathbf{p} \cdot \mathbf{x}} \right|^2.
 \label{fteq}
\end{equation}

In a small-angle approximation, the Fourier transform of $T$ can be related to spherical harmonics coefficients using $\tilde T(\mathbf{p})=a(\ell_p,m_p)$:
\begin{eqnarray}
 \notag \Delta^2(2 \pi \mathbf{p}) & = & 2 \pi |\mathbf{p}|^2 \, \left|a(\ell_p,m_p)\right|^2 \\
 \notag \Delta^2(\mathbf{p}) & = & \frac{|\mathbf{p}|^2}{2\pi} \, |a(\ell_p,m_p)|^2\\
 \Delta(\mathbf{p}) & = & |\mathbf{p}| \, \sqrt{ \frac{1}{2 \pi} |a(\ell_p,m_p)|^2 }. 
\end{eqnarray}
and thus, to express the power in the spherical harmonic degree corresponding to the normalized wave vector $\mathbf{p}$:
\begin{eqnarray}
 \Delta(\ell, m)
 & = & \hat{\ell} \sqrt{ \frac{1}{2 \pi} |a(\ell,m)|^2 }, 
\end{eqnarray} 
where $\hat{\ell} = \ell$ in the small angle approximation.  For any appreciable $\ell$, $\sqrt{\ell^2}~\approx~\sqrt{\ell (\ell+1)}$, and so
\begin{eqnarray}
 \Delta(\ell, m)
 & \rightarrow & \sqrt{ \frac{\ell(\ell+1)}{2 \pi} |a(\ell,m)|^2 }.
\end{eqnarray} 
From Eqs. (\ref{T1}) and (\ref{T2})
\begin{equation}
\left({\Delta T} \right)_\ell = \sqrt{ {{\ell(\ell+1)} \over {2\pi}} }  \,[a_{\ell,m}]
\label{uni}
\end{equation}
and so we see that
\begin{equation}
 [\Delta(\ell, m)] = [\left({\Delta T} \right)_\ell].
\end{equation}
So the power spectra produced in this analysis, and also in analyses typical in 21-cm cosmology, when computed directly with Eq.~(\ref{fteq}), are naturally in the $\left({\Delta T} \right)_\ell^2$ normalization.  Applying Eq.~(\ref{rt}) to Eq.~(\ref{uni}) we see that the physical units of $\Delta^2$ will indeed be squared temperature (e.g. K$^2$).


\bsp	
\label{lastpage}
\end{document}